\documentstyle[aas2pp4,psfig]{article}
\def\asca{{\it ASCA~}}
\def\rosat{{\it ROSAT~}}
\def\ginga{{\it Ginga~}}

\begin{document}
\title{X-ray Variability as a Probe of Advection-Dominated Accretion in
Low-Luminosity AGN}
\author{A. Ptak\altaffilmark{1}, T. Yaqoob, R. Mushotzky, P. Serlemitsos}
\authoraddr{ptak@astro.phys.cmu.edu; Carnegie Mellon University, 
Department of Physics, Pittsburgh, PA 15213}
\affil{NASA GSFC, Code 662, Greenbelt, MD 20771}
\author{R. Griffiths}
\affil{Carnegie Mellon University, 
Department of Physics, Pittsburgh, PA 15213}
\altaffiltext{1}{Current address: ptak@astro.phys.cmu.edu;
Carnegie Mellon University, 
Department of Physics, Pittsburgh, PA 15213}

\slugcomment{Submitted to ApJL, Mar. 2, 1998}
\slugcomment{Revised, Apr. 17, 1998}
\slugcomment{Accepted, Apr. 28, 1998}
\accepted{Apr. 28, 1998}
\lefthead{Ptak et al.}
\righthead{X-ray Variability from ADAFs in LLAGN}
\begin{abstract}
As a class, LINERs and Low-Luminosity AGN tend to show little or no
significant short-term variability (i.e., with time-scales less than a day).
This is a marked break for the trend of increased variability in Seyfert 1
galaxies with decreased luminosity.  We propose that this difference is due
to the lower accretion rate in LINERs and LLAGN which is probably causing
the accretion flow to be advection-dominated.  This results in a larger
characteristic size for the X-ray producing region than is the case in
``normal'' AGN.  Short-term variability may be caused by a localized
instability or occultation events, but we note that such events would
likely be accompanied by broad-band spectral changes.
Since the ADAF is more compact in a Kerr metric, it is
possible that the X-ray emission from ADAFs around rotating blackholes would be
more variable than X-ray emission from ADAFs in a Schwarzchild metric.
Similar variability arguments also apply to other wavelengths, and accordingly
multiwavelength monitoring of LLAGN could serve to ``map'' the ADAF regions.
\end{abstract}

 
\section{Introduction}
From the time LINERs were originally identified as a class (\cite{Heck80})
it has been debated whether or not they represent a low-luminosity version of
AGN (see \cite{fil96} for a review).  Recent optical work has shown that
$\sim 23\%$ of LINERs have broad $H\alpha$ lines (\cite{Ho97b}),
and X-ray observations of LINERs
show that the 2-10 keV spectrum is dominated by a power-law with an energy
index $\alpha \sim 0.8$ (\cite{Ptak97} and references therein).
These properties are highly suggestive that LINERs
are indeed AGN in many cases.  Recent surveys have shown that LINERs and
and other low-luminosity AGN (LLAGN; mostly Seyfert 2 galaxies) are very
common in the nuclei of nearby galaxies, so it important to understand these
objects.  Specifically, an important question is
to what extent do the properties of ``normal'' AGN
scale with luminosity or perhaps more precisely, what
properties scale with accretion rate.

Modeling of low-accretion rate blackhole systems suggests that when the
accretion rate drops below $\sim 0.1-0.01$ in 
Eddington units (i.e., $\dot{m}$
= $\dot{M}/\dot{M}_{Edd}$
, $\dot{M}_{Edd}$ = $1.4 \times 10^{18}\frac{M}{M_{\odot}}
\rm \  g \ s^{-1}$), 
the dominate mode of accretion will most likely 
be ``advection-dominated'' (\cite{abram95}, \cite{naryi95}).
The radiative efficiency of an 
advection-dominated accretion flow (ADAF) is very low, unlike the 
optically-thick, geometrically-thin, disks thought to be present in normal 
AGN where the accretion efficiency is probably on the order of 0.1.
In LLAGN and LINERs where the putative blackhole mass, $M_{BH}$, has been 
measured, $M_{BH}$ tends to be on the order of $10^{7-9}$ (c.f.,  
$M_{BH} \sim 3.5 \times 10^7 \rm M_{\odot}$ in NGC 4258, \cite{Green95};
$M_{BH} \sim 10^9 \rm M_{\odot}$ in NGC 4594, \cite{Kor96}).
Luminosities on the order of
$10^{40-41} \rm \ ergs \ s^{-1}$ therefore imply luminosities and accretion
rates 
on the order of $10^{-2}$ or less in Eddington units (assuming that on the
order of 10\% of the bolometric luminosity of these galaxies is
radiated in X-rays), well within the ADAF parameter space.  Indeed, the
presence of an ADAF in M81 (\cite{Petre93}) and NGC 4258 (\cite{L96})
has been suggested based in part on their low
accretion rates (see also \cite{mah97} and \cite{yibou98}).
However, we note that ADAF models for
accretion rates below $10^{-3}$ predict that the X-ray band emission would be
dominated by Bremsstrahlung emission from the $\sim 10^{9-10}$ K
electrons rather
than by inverse-Compton scattering of synchrotron
photons (c.f., \cite{naryi95}, \cite{yibou98}). 
With the exception of NGC 4258,
thermal Bremsstrahlung fits to the 2-10 keV emission have resulted in
temperatures on the order of $5-10 \times 10^7 \rm \ K$, ruling out $10^9$ K
emission.  Accordingly, the ADAF model predicts that 
the accretion rates in these
galaxies is probably not much below $\sim 10^{-3}$.
As with normal AGN, the X-rays produced from the ADAF arise from closer in to
the
AGN and are attenuated less by absorption than emission at other wavelengths.
In this letter we show that the X-ray temporal properties of LINERs and LLAGN
are more consistent with an ADAF than with a geometrically-thin disk.

\section{The Data}
Light curves were generated from \asca observations of a sample of
LLAGN and LINERs.  See
Ptak (1997) for a detail description of \asca and the analysis procedures.
Briefly, \asca consists of two sets of imaging spectrometers (the Solid-State
Imaging Spectrometers, hereafter referred to as the SIS, and the Gas Imaging
Spectrometer, hereafter referred to as the GIS).  \asca is sensitive in the
$\sim 0.4-10.0$ keV bandpass.
Several starburst galaxies are also included since they have spectra similar
to LINERs and LLAGN (see \cite{S96}, \cite{Ptak98}).
Here we discuss only the light curves
binned using the 2-10 keV data from the two GIS detectors.  The GIS are more
appropriate for timing analysis than the SIS since the GIS count rate is less
sensitive to aspect errors (note that this is usually only a concern for
4-ccdmode observations of weak sources).
The 2-10 keV bandpass was chosen because in most
of these galaxies the 0.4-2.0 keV bandpass is dominated by a soft component
that is probably thermal, is not necessarily associated with the AGN, and
may ``wash-out'' any variability in the hard power-law component.
\begin{figure}[h]
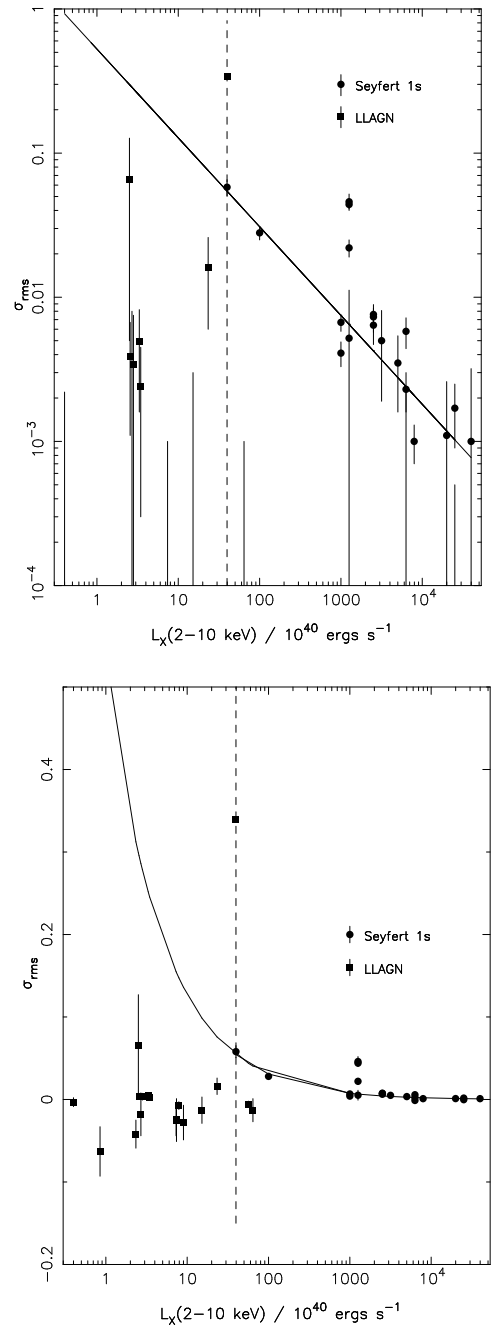

\hspace*{1cm}\psfig{figure=sy1_gal_sigrms_3.vps,height=8.5cm}
\vspace{5mm} \\
\hspace*{1cm}
\psfig{figure=sy1_gal_sigrms_lin_3.vps,height=8.5cm}
\caption{The ``variance'' of LLAGN galaxies (filled squares),
along
with the Seyfert 1 galaxies (filled circles) from Nandra et al. (1997),
plotted as a function
of 2-10 keV luminosity.  Note that several of the galaxies have been observed
by {\it ASCA} on multiple occasions.  The data are plotted with both a
logarithmic (top) and linear (bottom) vertical axis.
Note that some negative points are not displayed in the logarithmic plot.
The error bar for NGC
3079 is plotted with dashes to avoid confusion with NGC 4051.
The solid line
shows a simple power-law fit to the Seyfert 1 data.}
\end{figure}
\setcounter{figure}{0}

In order to characterize the variability (or lack of variability) observed in
each light curve, the parameter
\begin{equation}
\sigma_{rms}^2 = \frac{1}{N\mu^2}\sum_{i=1}^{N}[(X_i-\mu)^2-\sigma^2_i]
\end{equation}
was computed, hereafter referred to as ``variance''.  Here $N$ is the number of
light curve bins, $\mu$ is the mean count rate, $\sigma_i$ is the standard
deviation of the count rate in bin $i$, and $X_i$ is the count rate in bin $i$.
This quantity is listed
in Table 1 and plotted in Figure 1 as a
function of 2-10 keV luminosity.  Note that the variance computed from several
\asca observations of M51, M81, and NGC 3310 are presented, where the typical
separation of each observation was on the order of several months
(a more detailed
analysis of the multiple M81 observations will be presented in future work).
The motivation for including multiple observations, when available, is that
variance is observed to vary in Seyfert 1 galaxies (c.f., NGC 3227 in
\cite{george98}).
Also plotted are the variances
computed in Nandra et al. (1997) for Seyfert 1 galaxies, and it is obvious that
variance increases with decreasing luminosity in Seyfert 1
galaxies (see \cite{Nandra97}).  It is evident from this figure
that the same trend does not
extend down to the LINER and LLAGN galaxies.  In addition, simulations were
performed 
to ensure that the lack of variability was not due to the relatively
poor statistics of the LLAGN observations (see Ptak 1997).
Note that the variance observed in the starburst M82 is non-zero and
statistically similar to the variance computed from M81.  The implications of
these results are discussed below.

\section{Discussion}
For some of the galaxies in this sample, the lack of variability is probably
due to either the presence of multiple sources of the 2-10 keV emission or the
fact that the hard emission is scattered at distances greater than a light-day
from the nucleus.  The former case is most likely the situation for the
starburst galaxies, where while some of the hard flux may be due to
``hidden'' micro-AGN, much of hard flux may to be due to multiple
point-sources (supernovae and X-ray binaries) or hot gas (although as mentioned
above M82 is similar to M81 in its variance). The latter case is
most likely to be true for the Seyfert 2 galaxies NGC 3147, NGC 4258, and M51,
as suggested in \cite{Ptak96}.  However, note that in the case of NGC 4258,
the observed column density is on the order of $10^{23} \rm \ cm^{-2}$
which is consistent 
with the column densities in Seyfert 2 galaxies
observed by \ginga (\cite{Awaki92}) and \asca (\cite{turner97}).  While it
is possible that the hard X-ray flux from NGC 4258 is scattered around material
with a 
column on the order of $10^{24} \rm \ cm^{-2}$ or more, it is most likely that
the 2-10 keV flux we are observing is the direct nuclear continuum (note that
this argument is strengthened by the classification of NGC 4258 as a
Seyfert 1.9 in \cite{Ho97a}).

None of the above considerations are likely to be true for the brighter LINERs
in this sample, M81, NGC 3998, and NGC 4579 since each of these has been
observed to exhibit broad $H\alpha$ emission (\cite{Ho97b}),
making them ``Type-I'' LINERs,
analogues to the higher-luminosity Seyfert 1 galaxies.  In the cases of M81
(\cite{ishi96}), 
NGC 4579 (\cite{S96}) and, interestingly, the ``transition'' starburst-LINER
galaxy NGC 3628
(\cite{d95}; \cite{y95}), significant variability has been observed between
the \rosat and \asca observations, indicating that the nuclear sources
are dominating the emission.  It is therefore likely that in most of these
galaxies the dominant mode of accretion is fundamentally different from that
in Seyfert galaxies, since it is accretion that is driving the X-ray
luminosity and, by extension, the X-ray variability.

A fundamental difference between an ADAF and an optically-thick accretion disk
is that in an ADAF it is the flow itself that is producing the X-rays.  The
$\alpha$-disk solution predicts a temperature of
less than $10^{6}$ for $M_{BH} > 10^{7} \ \rm M_{\odot}$ (\cite{ss73}; see
\cite{frank92} for a review).  In this case (i.e., ``normal'' Seyfert
galaxies), the X-ray continuum is most likely produced by the inverse-Compton
scattering of UV photons from the ``cold'' accretion disk by energetic
electrons, possibly in a ``corona'' above the disk (c.f., \cite{HM91}).
In an ADAF, the
X-rays are produced by either the Comptonization of synchrotron
radiation by the electrons in the flow or by Bremsstrahlung emission from the
electron themselves.  In either case, the essence of the ADAF solution is that
the emission mechanism is inefficient (on accretion time-scales)
and a substantial volume contributes to
the X-ray emission.  Since the ADAF is likely to be nearly spherical,
most of the X-ray emission originates in a volume that is
probably a spherical annulus extending from $r \sim 5-10 \ R_{Schw}$ in the
case of a stationary blackhole or $r \sim 3-7$ in the case of a
maximally-rotating blackhole.  If
variability is due to a change in $\dot{m}$, then the {\it minimum} time for
the ADAF to respond is on the order of $\pi\bar{r}\gamma_g/c$, where
$\bar{r} = \frac{\int L_X(r)rdr}{\int L_X(r)dr}$,
$L_X(r)$ is the luminosity of the ADAF in the 2-10 keV
bandpass at $r$ and $\gamma_g$ is the gravitational time dilation
($\gamma_g \sim [1-R_{Schw}/r]^{-1/2}$).
$\bar{r}/c$ is probably on the order of 2 and 6 $R_{Schw}/c$
(i.e., near the inner-most stable orbits),
corresponding to $\sim 4$ and $11\frac{M_{BH}}{3.5 \times 10^{7} \rm \
M_{\odot}}$ ks light travel time-scales. 
More realistically, the time required for the ADAF to respond
to a change in $\dot{m}$ will depend on the sound speed, which approaches
$0.25c$ (stationary blackhole) or $0.5c$ (rotating blackhole) at small radii
(\cite{JK97}), resulting in a time-scale on the order of $\sim 6-20$ ks.  It is
evident that a Kerr metric should result in a more variable ADAF, while {\it
any} ADAF onto a blackhole at least as massive as $\sim 10^7 \rm \ M_{\odot}$
should not be variable on time-scales less than several to tens of ks.
The time-scales probed by the {\it ASCA} observation are on this order.

This analysis is based on the assumption that the any variability would be
caused by rapid
changes in $\dot{m}$ and that the ADAF will remain stable under these changes.
However, detailed modeling by \cite{Man96} and \cite{Tak97} showed that 
instabilities produced by rapid changes in $\dot{m}$ are likely to produce
``shots'' in the X-ray luminosity of an ADAF with time-scales on the order of 
several hundred $R_{Schw}/c$ (as observed in the low state of Cygnus X-1),
suggesting that the time-scales assumed above are very conservative.
It may be simply that the accretion rate itself is more steady in LLAGN than
it is in Seyfert galaxies.  Note that this also would then indicate a break
with
Seyferts, assuming that changes in accretion rate are ultimately responsible
for variability in Seyferts.  If occultation events are the cause of
variability in Seyferts, 
then the accretion flows in LLAGN lack material subtending a sufficient solid
angle to produce the same type of variability, again suggestive of a larger
region responsible for the production of X-rays.  Note also that if
variability on short time-scales is observed in a LLAGN believed to have an
ADAF is due to either an occultation event or an instability (where the X-ray
flux would be dominated by shocks rather than the steady ADAF flow emission),
then detectable spectral variability is likely to accompany the event.

Note that many of the arguments in this letter apply equally well to other
wavelengths, although with the exception of radio, the nuclear component of
the emission from a typical galaxy is absorbed, difficult to segregate from
extra-nuclear emission, or both.   However, since ADAF
models predict that the contribution of synchrotron cooling 
(dominating the radio) and Compton and bremsstrahlung cooling (dominating
in X-rays) to the ADAF luminosities varies only slightly as a function of
radius,
monitoring of the radio and X-ray luminosity of an ADAF should be further
test of ADAF models.  M81 was observed to vary in the radio on time-scales
of weeks (\cite{biet98}, \cite{Ho98}) and large-scale ($\Delta I/I \sim 1.7$) 
X-ray variability with a time-scale of
months and 20\% variability with time-scale of $\sim 1$ day was observed
by \asca (\cite{ishi96}, \cite{S96}).
Accordingly, the ADAF model predicts not only a specific broadband spectral
shape but also that the radio and X-ray flux from galaxies such as M81 would be
correlated over long periods of time (i.e., on time-scales sufficient for the
ADAF to be in equilibrium).  Finally, note that as the accretion rate
increases, the transition radius where the accretion flow changes from an ADAF
to a thin-disk flow would decrease (see \cite{Esin97}),
possibly resulting in short-term
variability as observed in Seyfert 1 galaxies.  In the case of hard X-ray
emission, this short-term variability should be accompanied by the spectral
features associated with $\alpha$-disks (i.e., Fe-K emission and the Compton
reflection ``hump'') which should not be present in AGN dominated by ADAFs.

\acknowledgements
The authors would like to thank Luis Ho for carefully reading an early version
of this manuscript and an anonymous referee for useful comments.

\newpage
\begin{deluxetable}{ccccc}
\tablecaption{X-ray Variability Data}
\tablehead{
\colhead{Galaxy} & \colhead{Mean$^a$} & \colhead{$\chi^2/dof^b$} &
\colhead{$L_{2-10 \rm \ keV}^c$} &
\colhead{$\sigma_{rms} \times 10^3$} \nl}
\startdata
NGC 253* & 0.11 & 29.9/31 & 0.40 & $-3.2 \pm 5.4$ \nl
NGC 3031 (M81) & 0.33 & 94.0/57 & 2.56 & $3.9 \pm 2.8$ \nl
NGC 3031 (M81) & 0.48 & 72.9/45 & 3.32 & $4.9 \pm 3.3$ \nl
NGC 3031 (M81) & 0.27 & 109/69 & 2.81 & $3.4 \pm 4.1$ \nl
NGC 3034 (M82)* & 0.57 & 31.4/18 & 3.46 & $2.4 \pm 2.1$ \nl
NGC 3079 & 0.0068 & 53.2/35 & 39.8 & $340 \pm 490$ \nl
NGC 3147 & 0.044 & 29.6/37 & 64.3 & $-13 \pm 14$ \nl
NGC 3310* & 0.037 & 16.2/16 & 7.30 & $-24 \pm 20$ \nl
NGC 3310* & 0.037 & 5.7/6 & 9.01 & $-28 \pm 21$ \nl
NGC 3310* & 0.037 & 14.3/15 & 7.39 & $-25 \pm 26$ \nl
NGC 3628 & 0.014 & 9.3/17 & 2.34 & $-42 \pm 17$ \nl
NGC 3998 & 0.17 & 47.9/57 & 56.8 & $-5.6 \pm 3.6$ \nl
NGC 4258 & 0.12 & 45.5/40 & 7.91 & $-7.4 \pm 4.7$ \nl
NGC 4579 & 0.090 & 92.5/56 & 23.3 & $16 \pm 10$ \nl
NGC 4594 & 0.062 & 29.0/24 & 15.2 & $-13 \pm 16$ \nl
NGC 5194 (M51) & 0.026 & 43.8/41 & 2.51 & $66 \pm 61 $ \nl
NGC 5194 (M51) & 0.027 & 28.2/31 & 2.70 & $-18 \pm 26$ \nl
NGC 6946* & 0.028 & 29.8/29 & 0.849 & $-63 \pm 30$ \nl
\tablenotetext{a}{Mean count rate in counts $\rm s^{-1}$.}
\tablenotetext{b}{$\chi^2$ resulting from the hypothesis of a constant
counting rate}
\tablenotetext{c}
{Observed 2-10 keV luminosity in units of $10^{40} \rm \ ergs \ s^{-1}$}
\tablenotetext{*}{Starburst galaxy (the optical classifications for all of the
galaxies in this sample is given in \cite{Ptak98})}
\enddata
\end{deluxetable}
\end{document}